\documentclass[twocolumn,showpacs,preprintnumbers,amsmath,amssymb]{revtex4}
\usepackage{graphicx}
\usepackage{dcolumn}
\usepackage{bm}
\begin{document}
\title{ Topological Properties of Integer Networks}
\author{Tao Zhou}
\author {Bing-Hong Wang}
\email{bhwang@ustc.edu.cn, Fax: +86-551-3603574.
 Correspondence Author}
\affiliation{%
Nonlinear Science Center and Department of Modern Physics,
University of Science and Technology of China, Hefei, 230026, PR
China
}%
\author {P. M. Hui}
\author {K. P. Chan}
\affiliation{%
Department of Physics, The Chinese University of Hong Kong, New
Territories, Shatin, Hong Kong, PR China
}%

\date{\today}

\begin{abstract}
Inspired by Pythagoras's belief that numbers represent the
reality, we study that the topological properties of networks of
composite numbers, in which the vertices represent the numbers
and two vertices are connected if and only if there exists a
divisibility relation between them.  The network has a fairly
large clustering coefficient of $\approx 0.34$, which is
insensitive to the size of the network. The average distance
between two nodes is shown to have an upper bound that is
independent of the size of the network, in contrast to the
behavior in small-world and ultra-small-world networks.  The
out-degree distribution is shown to follow a power law behavior
of the form $k^{-2}$.
\end{abstract}


\maketitle

Many social, biological, and communication systems form complex
networks, with vertices representing individuals or organizations
and edges representing the interactions between them
\cite{Review1,Review2,Review3}. Examples are numerous, including
the Internet, the World Wide Web, social networks of acquaintance
and other relationship between individuals, metabolic networks,
food webs, etc.
\cite{Pastor2001,Albert1999,Liljeros2001,Jeong2000,Stelling2002,Camacho2002}.
Empirical studies on real-life networks reveal some common
characteristics different from random networks and regular
networks.  Among these features, the most noticeable are the
small-world effect and scale-free
property\cite{WS1998,BA1999,BAJ1999}. In this Letter, inspired by
Pythagoras' belief that numbers represent absolute reality, we
study the topological properties of the most {\em natural} network
in Pythagoras' sense, namely the network consisting of integers.

The distance $d$ between two vertices in a network is defined as
the number of edges along the shortest path connecting them. The
average distance $L$ of the network is then defined as the mean
distance between two vertices, averaged over all pairs of
vertices. The average distance is one of the most important
properties in measuring the efficiency of communication networks.
In a store-forward computer network, for example, the most useful
measure characterizing the performance is the transmission delay
(or time delay) in sending a message through the network from the
source to the destination.  The time delay is approximately
proportional to the number of edges that a message must pass
through. Therefore, the average distance plays a significant role
in measuring the time delay.  Other important topological
properties include the clustering coefficient and degree
distribution.  The degree of a vertex $x$, denoted by $k(x)$, is
the number of the edges that are attached to the vertex. Through
the $k(x)$ edges, there are $k(x)$ vertices that are correlated
with or connected to the vertex $x$.  These neighboring vertices
form a set of neighbors $A(x)$ of $x$.  The clustering coefficient
$C(x)$ of the vertex $x$ is the ratio of the number of existing
edges among all the vertices in the set $A(x)$ to the total
possible number of edges connecting all vertices in $A(x)$.  The
clustering coefficient $C$ of the entire network is the average of
$C(x)$ over all $x$. Empirical studies indicate that many
real-life networks have much smaller average distances (with
$L\sim\ln S$, where $S$ is the number of vertices in the network)
than a completely regular networks and have much larger clustering
coefficients than the completely random networks. Studies on
real-life networks also show a power-law degree distribution of
the form $p(k)\sim k^{-\gamma}$, where $p(k)$ is the probability
density function for the degrees and $\gamma$ is an exponent. This
distribution falls off slower than an exponential, allowing for a
few vertices with large degrees. Networks with power-law degree
distribution are referred to as scale-free networks.

Here, we investigate a network in which the vertices represent a
set of positive integers. Two vertices $x$ and $y$ are linked by
an edge if and only if $x$ is divisible by $y$ or $y$ is a factor
of $x$. By definition, the topological structure of the network is
determined, once the set of vertices under consideration is known.
For example, Fig.1 shows the topological structure of a network
with the set of vertices representing the integers \{4, 6, 8, 9,
10, 12, 14, 15, 16, 18, 20, 21, 22, 24, 25, 26, 27, 28, 30\}.

Formally\cite{Bollobas2002}, a network is represented by a graph
$G(V,E)$, where $V$ is the set of vertices and $E$ is the set of
edges.  A graph $G$ is said to be connected if any two vertices of
$G$ are connected, i.e., one can go from one vertex to another
through the edges in the network. If a graph $G$ is not connected,
it consists of several disjoint components.  The example in Fig.1
has five components: $V1=$\{4, 6, 8, 9,10, 12, 14, 15, 16, 18, 20,
24, 27, 28, 30\}, $V2=$\{21\}, $V3=$\{22\}, $V4=$\{25\} and
$V5=$\{26\}. Although real-life networks are not always connected,
most previous studies have focused on connected graphs. In the
analysis of the network of integers, we only keep the largest
connected component if the given set of integers gives a
disconnected graph. With this choice, the sets of vertices \{4, 6,
8, 9, 10, 12, 14, 15, 16, 18, 20, 21, 22, 24, 27, 28, 30\} and
\{4, 6, 8, 9, 10, 12, 14, 15, 16, 18, 20, 24, 27, 28, 30\}, for
example, will generate the same graph consisting of 15 vertices
and 19 edges.

\begin{figure}
\scalebox{0.6}[0.55]{\includegraphics{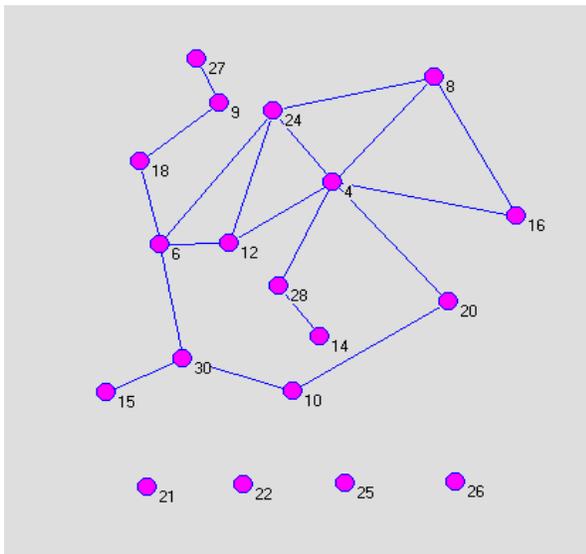}}
\caption{\label{fig:epsart} The topological structure of an
integer network with the set of vertices representing the integers
$V=$\{4, 6, 8, 9, 10, 12, 14, 15, 16, 18, 20, 21, 22, 24, 25, 26,
27, 28, 30\}. The corresponding graph consists of 19 vertices and
19 edges, and shows 5 disjoint components. The largest component
has 15 vertices and 19 edges, which is denoted by $G_{30}$.}
\end{figure}

Different sets of vertices (positive integers) will, in general,
lead to graphs of different properties.  For instance, the set of
integers $\{1,2,4,\cdots,2^{n}\}$ generates a complete graph, and
the set consisting of all the prime numbers and 1 leads to an
infinite star graph.  Here we focus on a special class of networks
of integers in which the vertices are {\em composite numbers}.  A
composite number is a positive integer greater than 1 and is not
prime.  We use the symbol $G_N$ to denote the network generated by
the set of composite numbers less than or equal to $N$, i.e.,
$V_N=\{x|x\in Q,4\leq x\leq N\}$, where $Q$ is the set of all the
composite numbers.  For example, the largest component of the
network in Fig.~1 is the graph $G_{30}$. Note that the number of
vertices in $G_N$ grows with $N$.  We have calculated the
clustering coefficients $C$ in the networks $G_{N}$ up to
$N=20,000$ and found that the clustering coefficients only
fluctuate very slightly about the value of $0.34$ with $N$ (see
inset of Fig.2).
\begin{figure}
\scalebox{0.8}[0.8]{\includegraphics{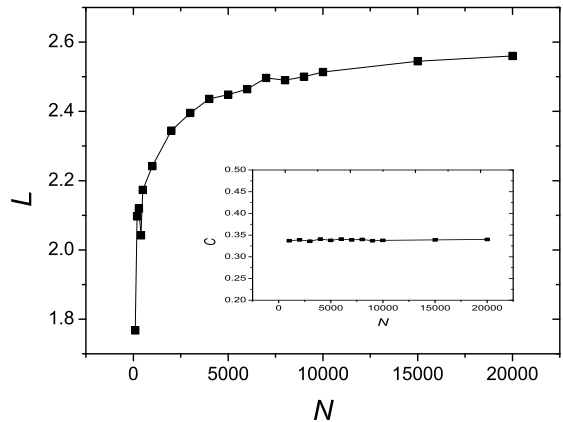}}
\caption{\label{fig:epsart} The average distance $L$ in the
network $G_{N}$ for network size up to $N=20,000$.  For $N>5,000$,
$L$ increases only slowly with $N$.  Analytically, there exists an
upper bound for $L$ that is independent of $N$.  The inset shows
the clustering coefficient $C$, which fluctuates only slightly
about the value 0.34 as $N$ increases.}
\end{figure}

\begin{figure}
\scalebox{1}[1.2]{\includegraphics{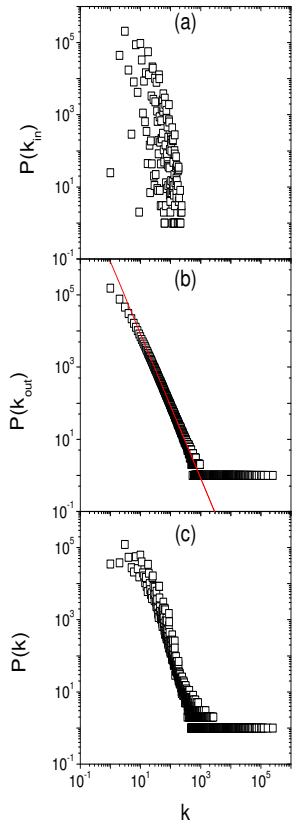}}
\caption{\label{fig:epsart} The (a) in-degree and (b) out-degree
distributions of the network $G_{1000000}$. The $y-axis$ gives
the total number of nodes having degree $k$.  The line in (b)
shows that the out-degree distribution follows a power law of the
form $\sim k^{-2}$. The total degree (with in and out degrees
counted together) distribution is shown in (c).}
\end{figure}

Figure 2 also shows the dependence of the average distance $L$ in
the networks $G_{N}$ for $N$ up to 20,000.  After a range of $N$
less than 5,000 for which $L$ increases from about 1.8 to 2.4, the
average distance increases fairly slowly with $N$ for $N > 5,000$.
Note that the average distance $L \sim \texttt{ln}N$ in
small-world networks\cite{WS1998} and $L \sim \texttt{lnln}N$ in
ultrasmall-world networks\cite{Rozenfeld2002}.  Here, we show that
the diameter\cite{ex} $D$ of $G_N$ must be smaller than a constant
$M$ that is independent of network size characterized by $N$,
i.e., $L<M$ in $G_{N}$ for arbitrary $N$, where M is a constant.
For any vertex $x\in G_N$, we first prove that the distance
between $x$ and the smallest integer 4 in $G_{N}$ satisfies
$d(x,4)\leq 4$. Since $x \in Q$, it can be written as
$x=p_1p_2\cdots p_q$, where $p_1\geq p_2\geq \cdots \geq p_q$ are
prime numbers.  If $q\geq4$, then $x/p_1p_2\in Q$, thus
$x/p_1p_2\in A(x)$.  Noting that $4x/p_1p_2\leq x\leq N$, there
exists a path of length 3 from $x$ to 4 through the vertices
$x/p_1p_2$ and $4x/p_1p_2$. If $q=3$ and $p_1>4$, analogously, a
path through the vertices $x/p_1$ and $4x/p_1$ exists between $x$
to 4.  If $q=3$ and $p_1<4$, then $x$ can only be $x$= 8, 12, 18,
and 27, for which 8 and 12 are connected directly to 4.  For
$N>107$, there exists a path from $x$ to $4$ through $4x$.
Therefore, for any vertex $x$ with at least 3 prime factors,
$d(x,4)\leq 3$. Finally, for $q=2$, since $A(x) \neq \varnothing$,
$y=2x$ must be in $A(x)$. Then $y$ has at least 3 prime factors:
$p_1$, $p_2$ and 2, leading to $d(y,4)\leq 3$ and thus $d(x,4)\leq
4$. Therefore, for any two vertices $a,b\in G_N$, we have
$d(a,b)\leq d(a,4)+d(b,4)\leq 8$. This implies $D\leq8$.  Since
the average distance $L<D$, the networks $G_{N}$ have a constant
upper bound for $L$.  The networks $G_{N}$ are, thus,
distinguished from the other networks by having a large clustering
coefficient $C$ ($\approx 0.34$) and a constant upper bound to the
average distance $L$.

Figure 3 shows the degree distributions of the network
$G_{1000000}$.  The degree of a vertex $x$ is the sum of the
in-degree and out-degree: $k(x)=k_{in}(x)+k_{out}(x)$, where the
in-degree of $x$ is defined as the number of elements in the set
$A(x)$ that are factors of $x$, and the out-degree of $x$ is the
number of elements in $A(x)$ that are divisible by $x$. Figure
3(a) and (b) show the in-degree and out-degree distributions. Note
that the range of in-degrees is smaller than that of out-degrees
in the network $G_{N}$.  The in-degree distribution does not show
any power law behavior. For the out-degrees,
$k_{out}(x)=\lfloor\frac{N}{x}\rfloor-1$. This leads to
$P(k_{out})\sim \frac{N}{(k+1)(k+2)} \sim k^{-2}$, for a range of
$k$ larger than unity.  This behavior is shown by the straight
line in Fig. 3(b).  The behavior for very large $k$ is limited by
the size of the network characterized by $N$. Counting the
in-degree and out-degree together, Fig.3(c) shows the total degree
distribution.

In summary, we proposed and studied a kind of networks that can be
regarded as the most natural, namely networks consisting of sets
of composite numbers as vertices.  Such networks exhibit
interesting features that are different from many previously
studied real-life networks.  Most noticeably, these networks show
a fairly large clustering coefficient. The average distance
between two vertices is shown to have an upper bound that is
independent of the network size.  The out-degree distribution
follows a power law.  The structure of such networks may be
useful, e.g., in communications, due to its special topological
properties. Mathematical objects, although unintentionally
defined to be so at the beginning, very often have close
connections to the physical world.  It is, therefore, worth
investigating the secret of other networks of mathematical
objects such as those consisting of groups and rings as vertices.

This work has been supported in part by the National Natural
Science Foundation of China under Grant No.10472116, No.70471033
and No.70271070, the Specialized Research Fund for the Doctoral
Program of Higher Education of China (SRFDP No.20020358009), and
the Foundation for Graduate Students of University of Science and
Technology of China under Grant No. KD200408. The authors thank
Miss Jun Liu for her assistance in preparing the manuscript. One
of us (P.M.H.) acknowledges the support of a Direct Grant for
Research at CUHK.


\begin{thebibliography}{Review1}
\bibitem{ex} The diameter $D$ of a network $G(V,E)$ is the maximal
distance: $D=\texttt{max}\{d(x,y)|x,y\in V\}$, where $d(x,y)$ is
the distance between two vertices $x$ and $y$.
\end{thebibliography}
\end{document}